\RequirePackage{lineno}

\documentclass[twocolumn,aps,prl,showpacs,superscriptaddress]{revtex4}
\usepackage{psfig}
\usepackage{graphicx}

\usepackage{lineno} 

\begin{document}

\setpagewiselinenumbers

\title{Studies of di-jet survival and surface emission bias in Au+Au collisions via angular correlations with respect to back-to-back leading hadrons}

\affiliation{Argonne National Laboratory, Argonne, Illinois 60439, USA}
\affiliation{Brookhaven National Laboratory, Upton, New York 11973, USA}
\affiliation{University of California, Berkeley, California 94720, USA}
\affiliation{University of California, Davis, California 95616, USA}
\affiliation{University of California, Los Angeles, California 90095, USA}
\affiliation{Universidade Estadual de Campinas, Sao Paulo, Brazil}
\affiliation{University of Illinois at Chicago, Chicago, Illinois 60607, USA}
\affiliation{Creighton University, Omaha, Nebraska 68178, USA}
\affiliation{Czech Technical University in Prague, FNSPE, Prague, 115 19, Czech Republic}
\affiliation{Nuclear Physics Institute AS CR, 250 68 \v{R}e\v{z}/Prague, Czech Republic}
\affiliation{University of Frankfurt, Frankfurt, Germany}
\affiliation{Institute of Physics, Bhubaneswar 751005, India}
\affiliation{Indian Institute of Technology, Mumbai, India}
\affiliation{Indiana University, Bloomington, Indiana 47408, USA}
\affiliation{Alikhanov Institute for Theoretical and Experimental Physics, Moscow, Russia}
\affiliation{University of Jammu, Jammu 180001, India}
\affiliation{Joint Institute for Nuclear Research, Dubna, 141 980, Russia}
\affiliation{Kent State University, Kent, Ohio 44242, USA}
\affiliation{University of Kentucky, Lexington, Kentucky, 40506-0055, USA}
\affiliation{Institute of Modern Physics, Lanzhou, China}
\affiliation{Lawrence Berkeley National Laboratory, Berkeley, California 94720, USA}
\affiliation{Massachusetts Institute of Technology, Cambridge, MA 02139-4307, USA}
\affiliation{Max-Planck-Institut f\"ur Physik, Munich, Germany}
\affiliation{Michigan State University, East Lansing, Michigan 48824, USA}
\affiliation{Moscow Engineering Physics Institute, Moscow Russia}
\affiliation{NIKHEF and Utrecht University, Amsterdam, The Netherlands}
\affiliation{Ohio State University, Columbus, Ohio 43210, USA}
\affiliation{Old Dominion University, Norfolk, VA, 23529, USA}
\affiliation{Panjab University, Chandigarh 160014, India}
\affiliation{Pennsylvania State University, University Park, Pennsylvania 16802, USA}
\affiliation{Institute of High Energy Physics, Protvino, Russia}
\affiliation{Purdue University, West Lafayette, Indiana 47907, USA}
\affiliation{Pusan National University, Pusan, Republic of Korea}
\affiliation{University of Rajasthan, Jaipur 302004, India}
\affiliation{Rice University, Houston, Texas 77251, USA}
\affiliation{Universidade de Sao Paulo, Sao Paulo, Brazil}
\affiliation{University of Science \& Technology of China, Hefei 230026, China}
\affiliation{Shandong University, Jinan, Shandong 250100, China}
\affiliation{Shanghai Institute of Applied Physics, Shanghai 201800, China}
\affiliation{SUBATECH, Nantes, France}
\affiliation{Texas A\&M University, College Station, Texas 77843, USA}
\affiliation{University of Texas, Austin, Texas 78712, USA}
\affiliation{University of Houston, Houston, TX, 77204, USA}
\affiliation{Tsinghua University, Beijing 100084, China}
\affiliation{United States Naval Academy, Annapolis, MD 21402, USA}
\affiliation{Valparaiso University, Valparaiso, Indiana 46383, USA}
\affiliation{Variable Energy Cyclotron Centre, Kolkata 700064, India}
\affiliation{Warsaw University of Technology, Warsaw, Poland}
\affiliation{University of Washington, Seattle, Washington 98195, USA}
\affiliation{Wayne State University, Detroit, Michigan 48201, USA}
\affiliation{Institute of Particle Physics, CCNU (HZNU), Wuhan 430079, China}
\affiliation{Yale University, New Haven, Connecticut 06520, USA}
\affiliation{University of Zagreb, Zagreb, HR-10002, Croatia}

\author{H.~Agakishiev}\affiliation{Joint Institute for Nuclear Research, Dubna, 141 980, Russia}
\author{M.~M.~Aggarwal}\affiliation{Panjab University, Chandigarh 160014, India}
\author{Z.~Ahammed}\affiliation{Lawrence Berkeley National Laboratory, Berkeley, California 94720, USA}
\author{A.~V.~Alakhverdyants}\affiliation{Joint Institute for Nuclear Research, Dubna, 141 980, Russia}
\author{I.~Alekseev~~}\affiliation{Alikhanov Institute for Theoretical and Experimental Physics, Moscow, Russia}
\author{J.~Alford}\affiliation{Kent State University, Kent, Ohio 44242, USA}
\author{B.~D.~Anderson}\affiliation{Kent State University, Kent, Ohio 44242, USA}
\author{C.~D.~Anson}\affiliation{Ohio State University, Columbus, Ohio 43210, USA}
\author{D.~Arkhipkin}\affiliation{Brookhaven National Laboratory, Upton, New York 11973, USA}
\author{G.~S.~Averichev}\affiliation{Joint Institute for Nuclear Research, Dubna, 141 980, Russia}
% ...

\collaboration{STAR Collaboration}\noaffiliation

\begin{abstract}
We report first results from an analysis based on a new multi-hadron correlation technique, exploring jet-medium interactions and di-jet surface emission bias at RHIC. Pairs of back-to-back high transverse momentum hadrons are used for triggers to study associated hadron distributions. In contrast with  two- and three-particle correlations with a single trigger with similar kinematic selections, the associated hadron distribution of both trigger sides reveals no modification in either relative pseudo-rapidity, $\Delta\eta$, or  relative azimuthal angle, $\Delta\phi$, from d+Au to central Au+Au collisions. 
We determine associated hadron yields and  spectra as well as production rates for such correlated back-to-back triggers  to gain  additional insights on medium properties. 

\end{abstract}

\pacs{25.75.-q}
                             
\maketitle


Di-hadron correlation measurements in heavy ion collisions have advanced the studies of hot and strongly interacting matter at RHIC.  The effects of jet-medium interactions are reflected  in the broad away-side distributions of associated hadrons in di-hadron azimuthal correlations and their softened transverse momentum ($p_T$) spectra~\cite{star_highpt,star_corr,phenix_highpt,phenix_corr}. For associated hadrons with transverse momentum between $1.0<p_T<2.0$~GeV/$c$ in addition to the overall broadening of the away-side  ($\Delta\phi\sim\pi$) a dip at $\pi$ is reported in~\cite{phenix_corr,MH,3pPRL}.  
Various scenarios and different mechanisms of jet-medium interactions have been proposed to explain the experimental observations~\cite{wii,wiii,wiv,wv}. 
On the near-side ($\Delta\phi\sim 0$) the 
modification of the correlation structure is also observed~\cite{star_corr,wvi,ridge,unT}. In~\cite{ridge} the same side peak is described by two components: the "ridge", a long-range $\Delta\eta$ plateau, and the small-$\Delta\eta$ jet-like peak. It is argued  that the ridge might arise from jet-medium interactions. However, recent  three-particle  analyses 
report possible decoupling of the jet and ridge features~\cite{3pRidge}.   
The existing three-particle correlation analyses do not 
specifically identify the away side jet hence the 
pseudo-rapidity correlation relative to the away-side jet cannot be studied.

Di-hadron azimuthal correlations of two high-$p_T$ hadrons have
been observed to exhibit jet-like peaks in both near- and away-sides~\cite{o31,phenix_nomodify}. Little modification is apparent in the shape of the peaks, but the amplitude of the away-side peak shows a strong suppression in central Au+Au collisions with respect to  d+Au collisions. 
The data may be interpreted in two scenarios: in-medium energy loss followed by in-vacuum fragmentation, and finite probability to escape the medium without interactions.
As jet-medium interactions manifest themselves via shape changes of correlation structures predominantly in the softer momentum range (below 2~GeV/$c$),
we investigate the modification dynamics using a new 3-particle ($2+1$) correlation technique, which measures the degree of correlation of a softer particle with a di-hadron trigger. The technique utilizes a second high-$p_T$ trigger on the away-side of the highest-$p_{T}$ particle in the event as an estimate  of the away-side jet axis direction.


The analysis presented in this letter used data collected by the STAR (Solenoidal Tracker at RHIC) experiment~\cite{STAR} in the years 2003-2004.
 Data samples consist of  5.5M d+Au, 14M minimum bias
Au+Au and 18M central triggered Au+Au events from collisions with the center-of-mass per nucleon pair energy of 200 GeV.  
Based on the charged track multiplicity at mid-rapidity ($|\eta|<0.5$) recorded by the STAR Time Projection Chamber (TPC)~\cite{TPC} the minimum bias Au+Au data are divided into 4 centrality bins. They correspond to 0-20\%, 20-40\%, 40-60\%, and 60-80\% of the total geometric cross-section.  The central-trigger data represent 
approximately 12\% of the total cross-section. To express centrality dependences we calculate via  Monte-Carlo Glauber model the number of participating nucleons ($N_{\text{part}}$) and the number of binary collisions ($N_{\text{coll}}$) for each centrality bin used in the analysis. 
The tracks selection for the analysis is identical to~\cite{MH}. 

We select the highest-$p_T$ particle in an event as the main (first) trigger and require $p_T^{\text{trig1}}$ to be between  5 to 10 GeV/$c$. The second trigger $p_T$, $p_T^{\text{trig2}}$, is selected to be above 4 GeV/$c$.
We consider two high-$p_T$ particles to be a di-jet-like pair if they conform to the requirement of relative azimuthal difference of $|\phi_{\text{trig1}}-\phi_{\text{trig2}}-\pi|<\alpha$.  The $\alpha$ is  chosen to be 0.2 corresponding approximately to 1$\sigma$ of the away-side peak width in relative azimuth for two high-$p_T$ hadrons~\cite{o31}.  
We construct the associated hadron distributions in $\Delta\phi$  and 
 $\Delta\eta$ coordinates with respect to each trigger direction.  The associated soft particle yield is measured relative to both triggers using the following expression 

\begin{linenomath*}
$$\frac{d^2N}{d\Delta\eta d\Delta\phi}=\frac{1}{N_{\text{trig}} \rm{\epsilon}_{\text{pair}}} \left(\frac{d^2N_{\text{raw}}}{d\Delta\eta d\Delta\phi}-a_{\text{zyam}} \frac{d^2N_{\text{Bg}}}{d\Delta\eta d\Delta\phi}\right)
$$
\end{linenomath*}
\noindent
where $N_{\text{trig}}$ is the number of trigger pairs, corrected for a dilution effect due to randomly associated pairs,  and ${d^2N_{\text{raw}}}/{d\Delta\eta\,d\Delta\phi}$ is the 
associated hadron distribution for all hadrons in triggered events. The associated hadrons $p_T$ (1.5~GeV/$c$~$<p_T^{\text{assoc}}<p_T^{\text{trig1}}$) is selected to coincide with the range where the away-side dip in di-hadron correlations is reported. The ``raw" distribution  is corrected for single-track efficiency and acceptance effects. The correction factor ($\rm{\epsilon}_{pair}$) includes single track reconstruction efficiency, which depends on $\eta$, $\phi$,  $p_T$ and multiplicity.  
It also corrects for the pair acceptance obtained by the mixed-event technique.
$\frac{d^2N_{\text{Bg}}}{d\Delta\eta d\Delta\phi}$ represents the total background, consisting of multiple terms.
The dominant background contribution is from combinatorics, and is accounted for by mixing trigger pairs and associated hadrons from different events, selected to have similar multiplicity and primary vertex position. The initially uniform distribution (apart from efficiency and acceptance effects) is then modulated by the flow term 
\begin{linenomath*}
$$f(\Delta\phi) = 1 + \frac{2 v_2^{\text{trig1}}v_2^{\text{assoc}}+2 v_2^{\text{trig2}}v_2^{\text{assoc}}\frac{\sin(2\alpha)}{2\alpha}}{1+2 v_2^{\text{trig1}}v_2^{\text{trig2}} \frac{\sin(2\alpha)}{2\alpha} } \cos(2\Delta\phi)$$
\end{linenomath*}
\noindent
where $v_2^{\text{trig1}}$, $v_2^{\text{trig2}}$, $v_2^{\text{assoc}}$ are sample-averaged values of the elliptical flow coefficient for primary and secondary triggers, and associated hadrons, respectively,
to account for elliptical flow effects, which we consider independent of $\eta$ in the range studied~\cite{flow,flowPHOBOS}. 
The multiplicity and $p_T$ dependence of the elliptical flow coefficients ($v_2$) for triggers  and associated  particles are obtained from averaging the results from the Event Plane and Four-particle Cumulant methods~\cite{flow}.  The background due to randomly associated triggers in the initial selection of the trigger pairs is also considered~\cite{3pPRL}.
 We  use di-hadron correlations for the appropriate kinematic selections to determine the shape of the related background distribution, and the signal-to-noise ratio from the trigger-trigger correlation to estimate the per-trigger contribution for such pairs.
The overall background level $a_{\text{zyam}}$ is estimated with the Zero-Yield at Minimum (ZYAM) method~\cite{zyam1,3pPRL,zyam2}.
For the 2+1 correlation we choose the zero-yield region to be more than 3$\sigma$
(corresponding to 1.3 radians) away from both jet-like peaks.
We also use a double Gaussian plus a $v_2$-modulated background fit to estimate the systematic error due to the ZYAM method. 
To obtain the transverse momentum spectra for associated hadrons we follow the same procedure 
 selecting the hadrons within
0.5~radians in relative azimuth and 0.5 in relative pseudorapidity of the respective trigger direction. 

Various sources of systematic uncertainties on measured yields are evaluated. 
Single track reconstruction efficiency contributes 
5\% overall normalization uncertainty for the associated hadron yields. This error does not affect same-side to away-side comparison for the correlated yields of the same data sample. 
The systematic uncertainty in the pair-acceptance correction due to finite statistics of the 
mixed-events is estimated to be less than 5\%.
The uncertainty due to the elliptical flow correction is estimated from the difference in the $v_2$ results from the Event-Plane and Cumulant methods. 
The uncertainty is found to be 5\% in most central, less than 1\% in most peripheral and 9\% in mid-central Au+Au events.
This uncertainty  is largely correlated between same- and away-side distributions, and is not applicable to d+Au events. 
Systematic uncertainty due to ZYAM normalization of the background level was estimated by varying  the $\Delta\phi$ range for minimum yield region, and is found to be $\sim$10\% in the Au+Au data and less than 5\% in d+Au events.  The related uncertainty is correlated between same- and away-sides. 
Systematic uncertainty due to correlated background subtraction from the di-jet sample is determined to be less than 3\% for both d+Au and Au+Au events. This error is correlated between the same- and away-sides, and was estimated by varying background normalization for the trigger-trigger correlation in a similar manner as for $2+1$ correlation.
Finally, the uncertainty in the di-hadron distributions used for estimating the correlated background term arises mainly from the  uncertainty in $v_2$ and ZYAM normalization. We evaluate these in a similar manner  and estimate the resulting effect to vary from less than 1\% in d+Au events to about 5\% in central Au+Au data. 

\begin{figure*}[hbt]
\centerline{\psfig{file=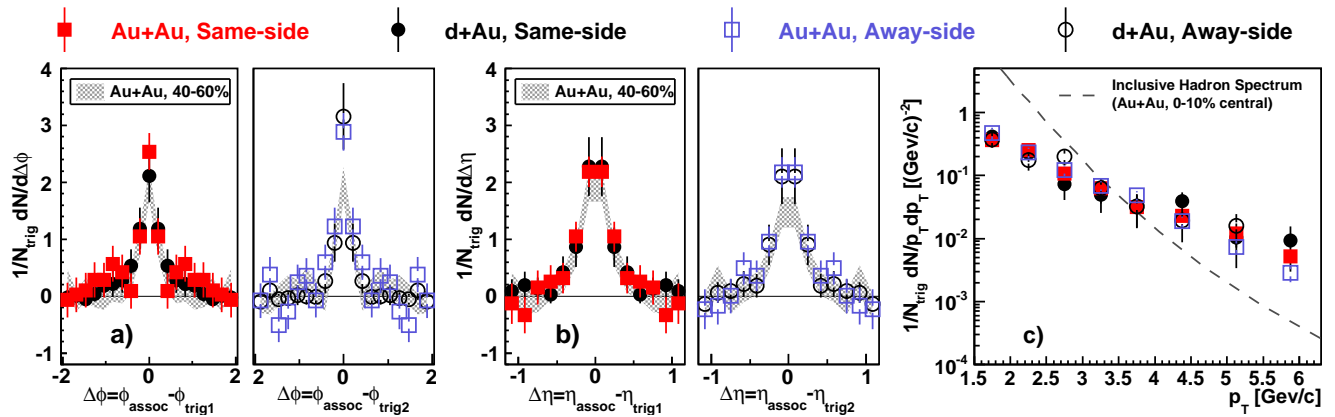,width=0.99\textwidth}}
\vspace*{-0.2cm}
\caption{
Projections of 2+1 correlation on  $\Delta\phi$ ($a$) and $\Delta\eta$ ($b$) 
 for 200 GeV top 12\% central-triggered  (squares) and mid-central (band) Au+Au  and minimum bias d+Au (circles) data.  Errors shown are statistical. The kinematic selection  is as follows: $5<p_T^{\text{trig1}}<10$~GeV/$c$, 4~GeV/$c$~$<p_T^{\text{trig2}}<p_T^{\text{trig1}}$, $1.5<p_T^{\text{assoc}}<p_T^{\text{trig1}}$.  
$c)$ Transverse momentum distributions per trigger pair for the same- and away-side hadrons associated with di-jet triggers ($|\Delta\phi|<0.5$, $|\Delta\eta|<0.5$). Errors are the quadrature sum of the statistical and systematic uncertainties.
Inclusive charged hadron distribution 10\% most central Au+Au data~\cite{star_raa} is shown for comparison. 
}
\label{fig:analysis}
\end{figure*}


Figure 1 shows $\Delta\phi$ and $\Delta\eta$ projections (symmetrized about 0) of the correlation function  defined above, measured for central and mid-central Au+Au collisions and compared with  the corresponding measurement from minimum bias d+Au data. The $\Delta\phi$ projection (Fig.~1$a$) reveals jet-like peaks on both the same- and away-sides.
The observed structures from  Au+Au data of all centralities  are consistent within errors  with the ones in d+Au collisions.  
This constitutes the first observation
of  the away-side correlation structure peaked in both $\Delta\phi$ and $\Delta\eta$ for the associated hadrons in this kinematic range. 
These results differ significantly from earlier di-hadron correlation observations, where substantial  differences in both yields and shapes on both the same and away-sides were observed between Au+Au and d+Au data~\cite{MH}.  
We observe no evidence of the Mach-cone effect in the three particle (2+1) correlations studied.
We attempt to explore the ridge properties by looking at $\Delta\eta$ projections for di-jet triggered correlations for same-side as well as away-side triggers (Fig.~1$b$). We find that the same-side integrated yields at $0.5<\Delta\eta<1.5$ are consistent with zero; however, statistical limitations ban exclusion of the ridge magnitudes reported in~\cite{ridge}.  
The $2+1$ results provide no evidence of a long-range "ridge" associated with the direction of either of the  high-$p_T$ triggers, consistent with lack of medium-induced effects on those di-jets selected by this analysis. 
The possible effects of radial flow on Mach-cone-like and ridge signals
 in di-hadron correlations have been studied in~\cite{flow-effects}. We expect no
such radial flow-related effects for tangentially emitted trigger pairs.

We further explore the associated hadron spectra to complement the correlation shape analysis. Figure~1$c$ shows  same- and away-side associated hadron yields plotted vs transverse momentum in d+Au and 12\% central Au+Au collisions.
Within the uncertainties no difference is observed between associated hadron spectra on same-side or away-side,  Au+Au or d+Au data samples.  Earlier measurements from di-hadron correlations in a similar kinematic range have shown significant softening of the away-side associated hadron spectra~\cite{star_corr}.

To check for the possible away-side softening which is indicative of energy
deposition in the medium, we have estimated the jet energy difference between same- and away- sides.
We sum the 
transverse momentum of trigger particle and the charged hadrons used for the spectra as an estimate for jet energy in the fiducial range. 
We then calculate the difference between such energy estimates for the same-side and away-side for each of the data samples. 
The direct subtraction of the momentum-weighted same-side and away-side distributions used for this calculation results in a major improvement of systematic uncertainties as no background re-normalization/subtraction  is needed and uncertainties on elliptical flow cancel. 
We find that in the 12\% central Au+Au data 
$\Delta(\text{Au+Au}) = \Sigma(p_T)^{\text{same}}-\Sigma(p_T)^{\text{away}}=1.59\pm 0.19$~GeV/$c$, similar to the minimum bias d+Au data value of $\Delta(d+Au) = 1.65 \pm 0.39$~GeV/$c$. 
The initial state kinematic effects are expected to cause similar differences~\cite{renk_kt,renk_jet}, and  calculated theoretically to be 1.6~GeV/$c$.

\begin{figure*}[thb]
\centerline{\psfig{file=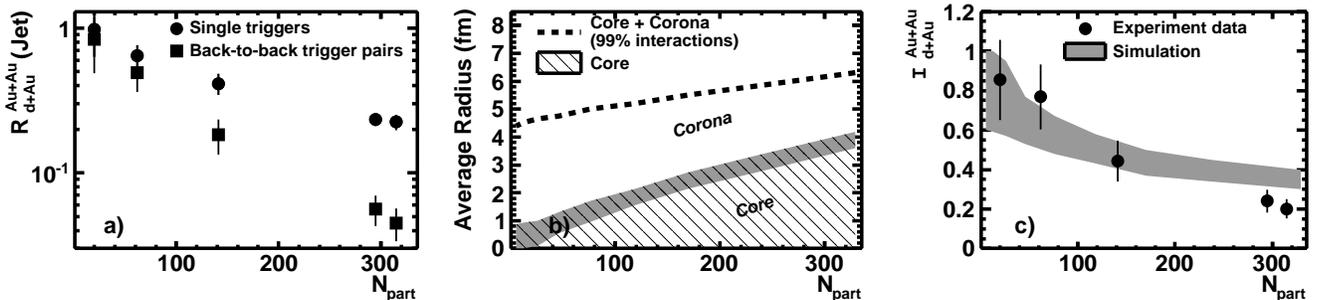,width=0.99\textwidth}}
\vspace*{-0.5cm}
\caption{
$a)$ Relative production rates for jets and di-jets in Au+Au collisions of different centralities with respect to d+Au data. $b)$ Calculations of the Glauber-based core/corona  model that accommodates inclusive hadron suppression and single jet production rate results.  
$c)$ Conditional di-jet survival probability in Au+Au data compared to d+Au reference. The band shows the expectations from Glauber-based core/corona model described in the text. Error bars in $a)$ and $c)$ are the quadrature sum of the statistical and systematic uncertainties.} 
\end{figure*}
\label{fig03}

The absence of a jet quenching signal or a suppression
of associated hadrons suggests that the $2+1$ particle triplets considered in this analysis are biased towards surface jet emission, i.e. di-jet events where both jets are emitted nearly tangentially to the medium surface. 
In such a scenario, one would expect that di-jet production rates are determined by the surface area of the fireball. We investigate this possibility by studying the centrality dependence of jet and di-jet production rates in 200 GeV Au+Au data.

Figure 2$a$ shows the centrality dependence of the nuclear modification factors $R_{\text{d+Au}}^{\text{Au+Au}}$ (ratio of binary-scaled per-event trigger counts in Au+Au and d+Au data) for the primary  (single) triggers and di-jet triggers.
The observed $R_{\text{d+Au}}^{\text{Au+Au}}$ for single triggers is consistent with the suppression factors observed in inclusive charged hadron measurements~\cite{star_raa}, while correlated trigger pair production rates are suppressed even further. 

To examine if these results are consistent with the purely tangential jet production scenario, 
 Monte Carlo Glauber calculations to find the spatial distribution of hard scattering were performed for each centrality bin. 
We assume a simplistic scenario  where the interaction zone in heavy-ion collisions  consists of a completely opaque core (full jet attenuation) surrounded by a  permeable corona (no jet-medium interactions).
Energy loss fluctuations as discussed in~\cite{RenkM} have not been taken into account.
Participant eccentricity for the medium core's shape has been used. This has been found most suitable for elliptical flow results~\cite{g33}.

The size of the core is estimated directly from the $R_{\text{AA}}$ suppression measurements~\cite{phenix_raa,star_raa} and the calculated core eccentricity.
The total size of the interaction area is calculated in a similar way, requiring the integrated collision density inside this area to be 99\% of the total $N_{\text{coll}}$.
We further assume that all observed trigger hadrons come from the corona. Core/corona sizes that accommodate the inclusive hadron suppression results are shown in Fig.~2$b$. The gray band shows propagation of uncertainties in the published $R_{\text{AA}}$ values into our calculations. Figure 2$c$ displays the  conditional di-jet survival rates.
The expected  rates are shown  as a band, where the width reflects the uncertainty in the published $R_{\text{AA}}$ data.

To estimate the conditional di-jet survival probability in the data, we use the double ratio $I^{\text{Au+Au}}_{\text{d+Au}}=\frac{R_{\text{AA}} (\text{trigger pairs})} {R_{\text{AA}}(\text{single triggers})}$,            
which reflects any changes in probability to find an away-side trigger for each primary trigger in Au+Au data relative to d+Au. 
For 12\% most central Au+Au data the likeliness of finding an away-side trigger for each primary trigger, which might be interpreted as conditional di-jet survival probability, is approximately $20 $ \% $ \pm 2 \text{(stat)} \pm 4.5 \text{(syst)} $ compared to d+Au. 
This ratio is shown as symbols in Fig. 2$c$ for comparison with the above model.
However, we must point out that core emission where neither of the di-jets
interacted with the medium cannot be ruled out by this analysis.

To summarize: we attempt to investigate mechanisms of jet-medium interactions in ultra-relativistic heavy ion collisions using a novel technique involving three particle ($2+1$) correlations of hadrons associated with a correlated pair of back-to-back high $p_{T}$ particles.
Both same-side and away-side peaks of di-jet triggered correlations from central Au+Au data are found similar to the structures observed in d+Au  collisions at the same energy. We observe the same result in all centrality bins of minimum bias Au+Au data. We measure associated hadron spectra on each side of a di-jet primary trigger, and find that spectral shapes and integrated yields in central Au+Au data are similar on the same- and away -sides, and are consistent with those measured in d+Au data. 
In contrast with earlier measurements with a single trigger,  we observe no modification in correlations of associated hadrons in the momentum range of $p_T^{\text{assoc}}>1.5$~GeV/$c$ with the
kinematic selection studied.
Systematic assessment of di-jet production rates supports tangential emission bias in a simplistic core/corona scenario, showing no evidence for di-jets interacting with the core. %

We thank the RHIC Operations Group and RCF at BNL, the NERSC Center at LBNL and the Open Science Grid consortium for providing resources and support. This work was supported in part by the Offices of NP and HEP within the U.S. DOE Office of Science, the U.S. NSF, the Sloan Foundation, the DFG cluster of excellence `Origin and Structure of the Universe'of Germany, CNRS/IN2P3, FAPESP CNPq of Brazil, Ministry of Ed. and Sci. of the Russian Federation, NNSFC, CAS, MoST, and MoE of China, GA and MSMT of the Czech Republic, FOM and NWO of the Netherlands, DAE, DST, and CSIR of India, Polish Ministry of Sci. and Higher Ed., Korea Research Foundation, Ministry of Sci., Ed. and Sports of the Rep. Of Croatia, and RosAtom of Russia.

{}

\end{document}